\documentclass[a4paper]{PoS_teddy}

\def\apj{{\it Astrophys.~J.}}
\def\apjl{{\it Astrophys.~J.~Lett.}}
\def\apjs{{\it Astrophys.~J.~Supp.}}

\def\araa{{\it Ann.~Rev.~Astron.~Astrophys.}}

\def\lrr{{\it Liv.~Rev.~Rel.}}

\def\prl{{\it Phys.~Rev.~Lett.}}

\title{The MERger-event Gamma-Ray (MERGR) Telescope}

\ShortTitle{MERGR Telescope}

\author{\speaker{L. J. Mitchell\thanks{Development of MERGR at NRL is supported by the Chief of Naval Research.}}, J. E. Grove\thanks{Presenter.}, B. F. Phlips, C. C. Cheung, M. Kerr, R. S. Woolf\\
        Naval Research Laboratory\\
        E-mail: \email{Lee.Mitchell@nrl.navy.mil, Eric.Grove@nrl.navy.mil}}

\author{M. S. Briggs\\
        University of Alabama, Huntsville}

\author{J. S. Perkins\\
        NASA Goddard Space Flight Center}

\abstract{
We describe the MERger-event Gamma-Ray (MERGR) Telescope intended for deployment by $\sim$2021. MERGR will cover from 20 keV to 2 MeV with a wide field of view (6 sr) using nineteen gamma-ray detectors arranged on a section of a sphere. The telescope will work as a standalone system or as part of a network of sensors, to increase by $\sim50\%$ the current sky coverage to detect short Gamma-Ray Burst (SGRB) counterparts to neutron-star binary mergers within the $\sim$200 Mpc range of gravitational wave detectors in the early 2020's. Inflight software will provide realtime burst detections with mean localization uncertainties of 6$^{\circ}$ for a photon fluence of 5 ph cm$^{-2}$ (the mean fluence of Fermi-GBM SGRBs) and $<3^{\circ}$ for the brightest $\sim5\%$ of SGRBs to enable rapid multi-wavelength follow-up to identify a host galaxy and its redshift. To minimize cost and time to first light, MERGR is directly derived from demonstrators designed and built at NRL for the DoD Space Test Program (STP). We argue that the deployment of a network that provides all-sky coverage for SGRB detection is of immediate urgency to the multi-messenger astrophysics community.
}

\FullConference{7th Fermi Symposium 2017\\
		15-20 October 2017\\
		Garmisch-Partenkirchen, Germany}

\begin{document}

\section{Introduction}
\vspace{-0.15in}

Gamma-Ray Bursts (GRBs), the most luminous transient events in the Universe, were discovered during the Vela satellite programs in the 1960s.  They were subsequently determined to be of cosmic origin based on their individual directions \cite{kle73,cli73} and extragalactic in nature because of their isotropy \cite{pac99}. GRBs can observationally be separated into two classes \cite{kou93} based on durations of $<$2s (short GRBs) and $>$2s (long GRBs). After decades of study, the emergent picture is that these sub-classes originate from compact binary mergers and the collapse of massive stars, respectively. 

The Advanced LIGO detections of the first gravitational wave (GW) sources in 2015 have opened a new multi-messenger era of astrophysics. These initial GW detections were somewhat unexpected because they resulted from the mergers of black hole (BH) binaries with individual masses of up to $\sim$30 solar masses, challenging binary evolution scenarios.  While the number of BH-BH merger detections by Advanced LIGO/Virgo is increasing \cite{abb17b}, such events are not expected to produce electromagnetic counterparts (but for GW150914, see \cite{con16} and Kocevski et al., these proceedings). 

Importantly, neutron-star (NS) binary mergers have a long theoretical basis that they should produce short GRBs \cite{ber14} and the recent LIGO/Virgo-correlated Fermi GBM detection of GRB170817A confirmed this for the case of a $\sim$40 Mpc distant NS-NS merger, GW170817 \cite{abb17,abb17c}. The most prolific GRB detectors are currently the Swift BAT and Fermi GBM because of their sensitivity and wide field of view (FoV). These missions are already sensitive to the crop of NS-binaries that are expected to be detectable soon in by GW observatories. Fermi GBM, which observes the entire unocculted sky, has demonstrated particular efficiency at detecting SGRBs because of its wide FoV \cite{bur16}.  While successful, these missions have been active for many years, and complementing or replacing their capability is an important consideration for future GW followup.

MERGR's primary science objective is the detection and localization of short GRBs ($<$2s) from mergers of NS-NS binaries or possibly also NS-BH systems. The instrument is designed to complement existing (Fermi GBM and Swift BAT) and future GRB detection systems, such as BurstCube \cite{rac17}, to provide all-sky coverage and improved localization of such events. Of immediate importance are the NS-binary systems within the gravitational-wave detection range of $\sim$200 Mpc expected from the planned upgrades to the Advanced LIGO-Virgo detectors in the $\sim$2019--2023+ timeframe \cite{abb16}.

\begin{figure}[t]
\centering
\includegraphics[height=0.25\textheight]{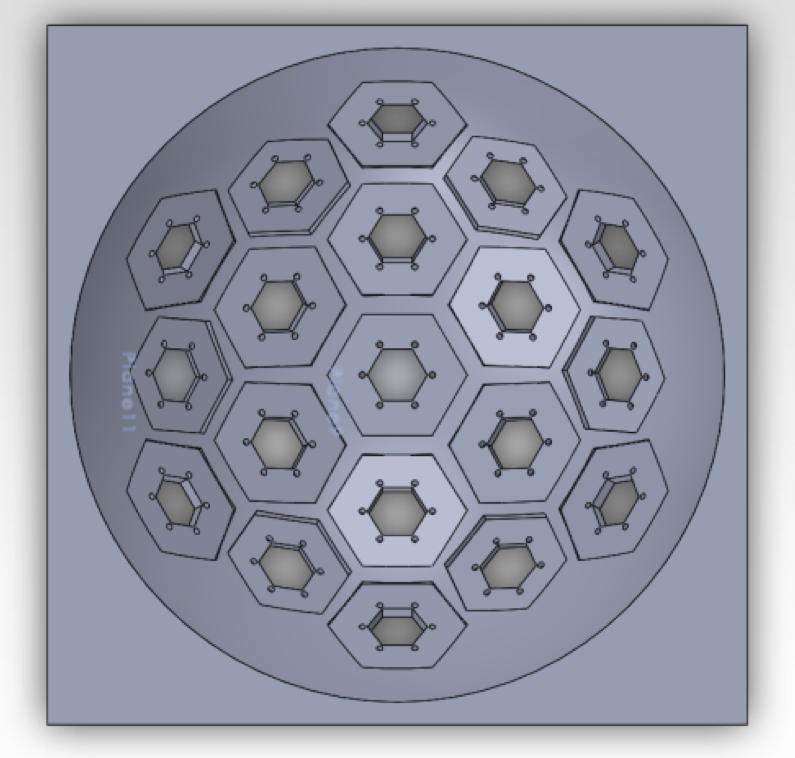}\includegraphics[height=0.25\textheight]{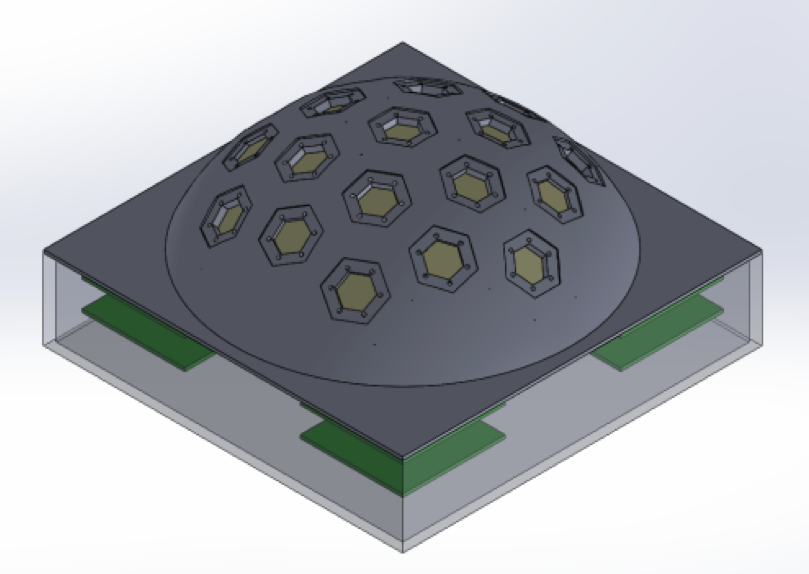}
     \caption{(Left) View of MERGR from the instrument's zenith showing the locations of the 19 hexagonal detectors in a closely-packed dome geometry. (Right) Off-axis view showing MERGR's electronics bay.}
     \label{fig1}
\end{figure}

\begin{figure}[b]
\begin{minipage}[c]{0.46\textwidth}
\includegraphics[width=\textwidth]{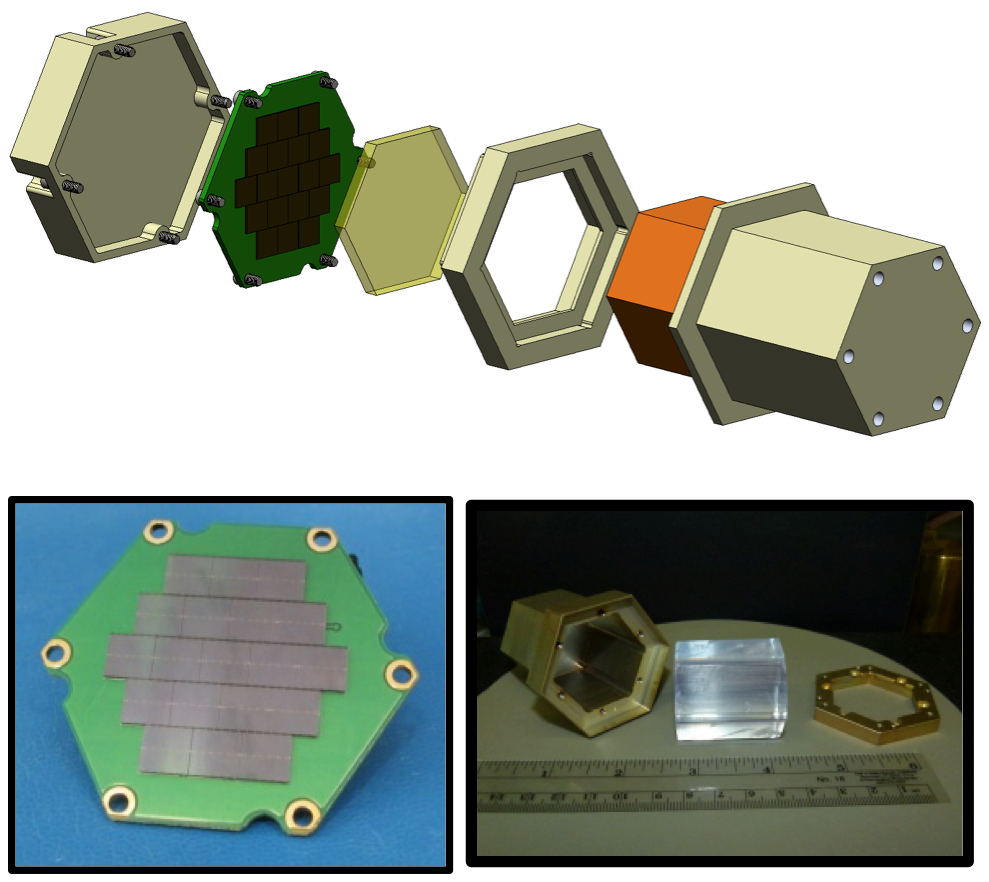}
\end{minipage}\hfill
\begin{minipage}[c]{0.49\textwidth}
     \caption{SrI$_{\rm 2}$:Eu detectors developed for use on SIRI-2. The MERGR detectors would be identical, except for the thickness (1.3 cm). 
(Top) Exploded CAD rendering of detector assembly showing the individual components of the detector. 
(Bottom Left) SiPM array developed by NRL for read out of the hexagonal crystals uses 19 SensL SeriesJ SiPM. 
(Bottom Right) SIRI-2 flight model hexagonal crystal prior to installation into detector housing.}
     \label{fig2}
\end{minipage}
%\vspace{-4mm}
\end{figure}

\section{Instrument Description}
\vspace{-0.15in}

A low-cost GRB localizer can be designed and built using mature technologies.  MERGR is designed to cover the energy range from 20 keV to 2 MeV with a wide FoV (6 sr) and good flux sensitivity to maximize detection probability for fast transients.  The instrument's instantaneous sky coverage is approximately $50\%$. This configuration seeks to achieve the best angular resolution while minimizing the loss in effective area. The segmented design of MERGR allows reconstruction of the burst location with sufficient accuracy to seed more precise follow-on searches for counterparts in other wavebands. MERGR relies extensively on experience and heritage from the Strontium Iodide Radiation Instrumentation series of gamma-ray instruments we built (SIRI-1) and are currently building (SIRI-2) for launch by the DoD Space Test Program (STP). The DoD STP provides a launch opportunity and funds for its integration into the spacecraft bus, launch services, and the first year of experiment data collection. Using existing algorithms developed and tested at NRL, instrument flight software will provide real-time burst detections and locations, which will be transmitted to the Gamma-ray Coordinates Network for worldwide distribution.

Figure 1 shows a CAD rendering of the MERGR instrument with a detector array consisting of 19 hexagonal close-packed SrI$_{\rm 2}$:Eu scintillators, each with a crystal diameter of 3.8 cm and a thickness of 1.3 cm. SrI$_{\rm 2}$ has high density (4.2 g cm$^{-3}$) and good stopping power.  It has high light output and exceptionally good energy resolution for a scintillator ($\sim4\%$ at 662 keV).  While the resolution provides no particular benefit for the smooth gamma-ray spectra of GRBs, it has no disadvantages and allows for the study of phenomena requiring spectroscopy. The detectors are mounted on a dome that is the segment of a sphere with radius 18.3 cm. This configuration provides a wide FoV, good sensitivity, and good source localization.  The center of the detector is at $0^{\circ}$ (with respect to the instrument zenith). An inner ring of six detectors is at $22.5^{\circ}$.  An outer ring alternates with six detectors at $39^{\circ}$ and six detectors at $45^{\circ}$. 

Located just under the detector array are the data acquisition system and satellite bus interfaces, shown in Figure 1. The scintillation light is read out by a NRL-developed custom array of 19 6-mm SensL SeriesJ SiPMs (Figure 2).  The overall instrument has a length of 35.6 cm, width of 35.6 cm, and height of 19 cm. The estimated mass of the telescope is 12.7 kg. Nominal power consumption is estimated to be 22 W.

\begin{figure}[t]
\centering
\includegraphics[height=0.252\textheight]{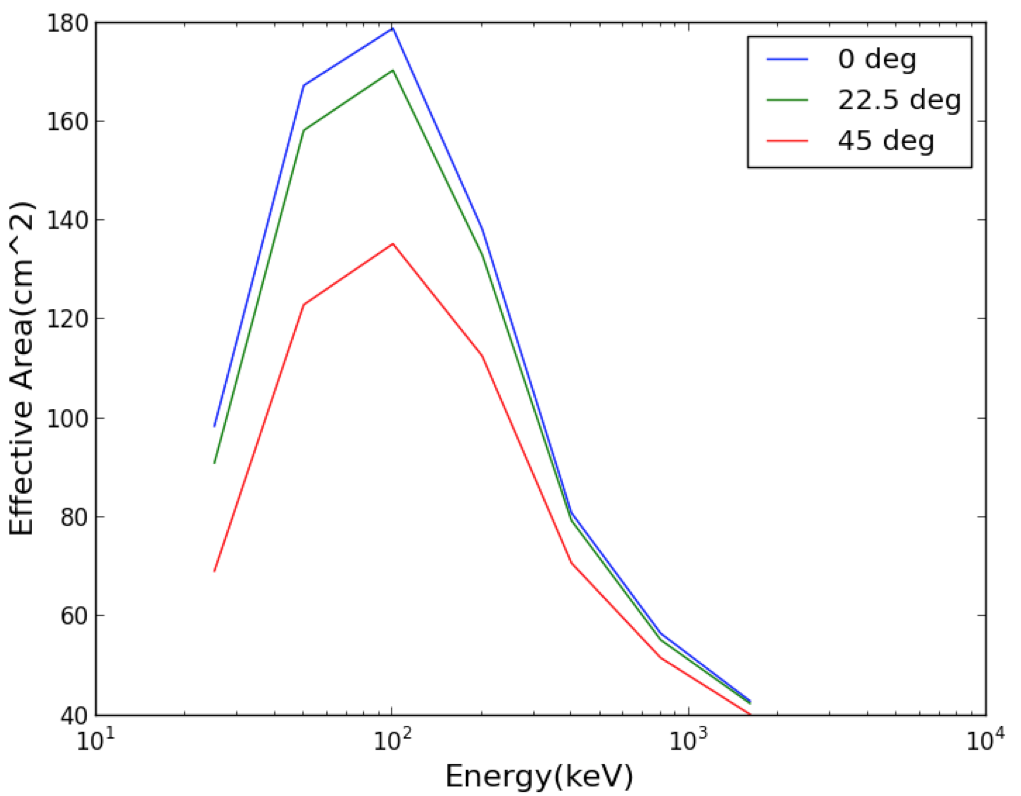}\includegraphics[height=0.25\textheight]{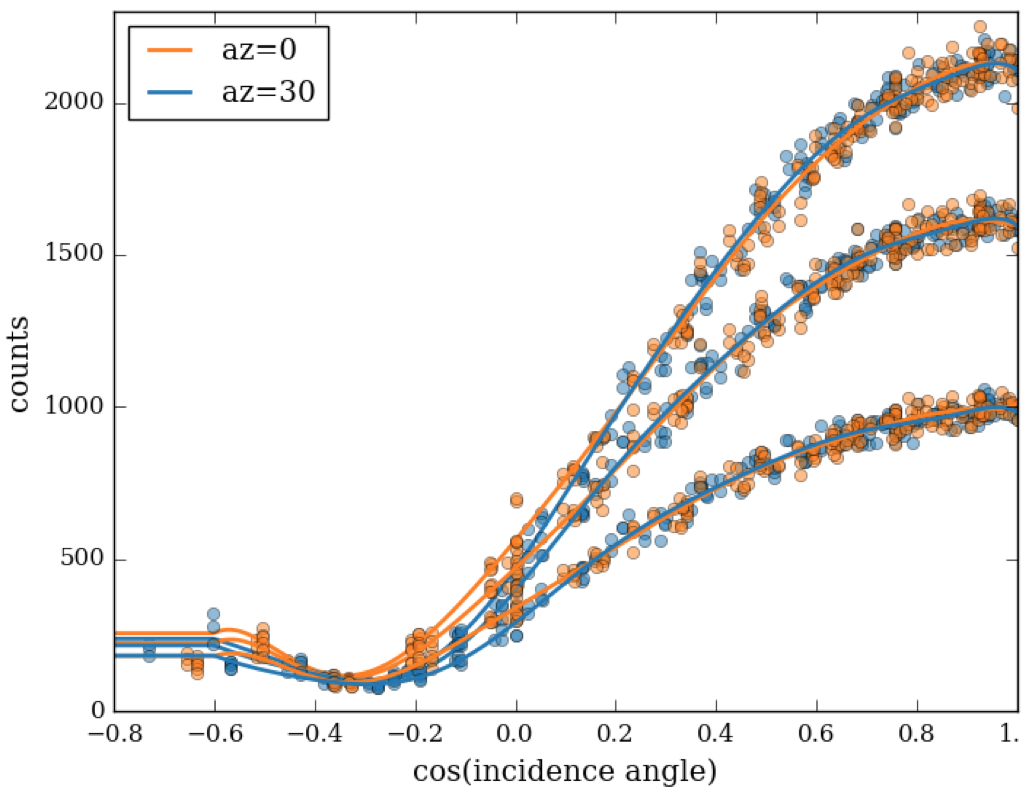}
     \caption{(Left) MERGR effective area for three viewing angles relative to the instrument boresight.
(Right) Observed counts a function of incidence angle for the two primary symmetries of the detector (face and facet).  The angular dependence can be fit with a simple polynomial.  The three clusters show thresholds at 20, 50, and 100 keV (from top to bottom).}
     \label{fig3}
\end{figure}

\section{Instrument Performance}
\vspace{-0.15in}

Using Monte Carlo techniques, we determined the effective area as a function of energy for three angles from the instrument boresight, $0^{\circ}$, $22.5^{\circ}$, and $45^{\circ}$ (Figure 3, Left).

We next considered the response of the detector over a fine-grained range of incidence angles and azimuth to determine the angular response of each element of the detector, including attenuation of soft photons in the aluminum enclosures.  Simulations show the response of each detector depends almost entirely on the incidence angle, and modestly on the azimuth angle (az), reflecting the small asymmetry in the hexagonal crystals (Figure 3, Right).  By fitting and applying this response, we are able to construct a simple but accurate geometric model of the detector, which comparison to the Monte Carlo simulations shows better than $10\%$ accuracy over the full range of incidence angles.

Using this model, we can apply maximum likelihood methods to evaluate the detector response to a wide range of burst position and fluence.  To characterize the localization performance, we evaluated the Fisher information, which is the expectation of the second derivative of the log Poisson likelihood with respect to the model parameters, specifically fluence and source position.  The Fisher information may be inverted to estimate the typical covariance matrix of the model parameters, and from this their expected precision.  We performed this exercise for three energy thresholds (20, 50, and 100 keV) both without background and with a background estimated from GBM counts selected at the appropriate threshold and scaled to the MERGR crystal volumes and a typical burst length of 0.3 s.  To derive a typical burst localization performance, we take the geometric mean of the two angular precisions and average them over the sky.  These results appear in Figure 4 (Left), which shows that a 20 keV threshold is optimal and that the current design delivers a performance given approximately by $r_{\rm 68} = 5.9^{\circ}$  (fluence / 5 ph cm$^{-2}$)$^{-0.7}$.

Finally, we considered burst sensitivity by simulating events and then performing maximum likelihood fits in position and fluence.  By computing the difference in log likelihoods of two models (burst and no burst) we obtain a test statistic, or TS, for the presence of a burst. We first simulated 30,000 background-only cases to determine the distribution of TS in the null hypothesis.  The distribution is approximately exponential, and values of 10 or larger occur only a few times.  If we consider a 0.3 s trigger window, a threshold of 10 then corresponds to about 50 triggers per day, and consequently any burst above this threshold should be able to be recorded and downlinked with a modest telemetry budget.  We then simulated bursts at three incidence angles and a range of fluences to determine the fraction of bursts above this threshold.  Unlike localization, we are more sensitive to the presence of bursts that occur on-axis as the effective area is highest (Figure 4, Right), indicating we are complete above a fluence of 1.0 ph cm$^{-2}$ for on-axis bursts and 1.5 ph cm$^{-2}$ for off-axis bursts.

\begin{figure}[t]
\centering
\includegraphics[height=0.25\textheight]{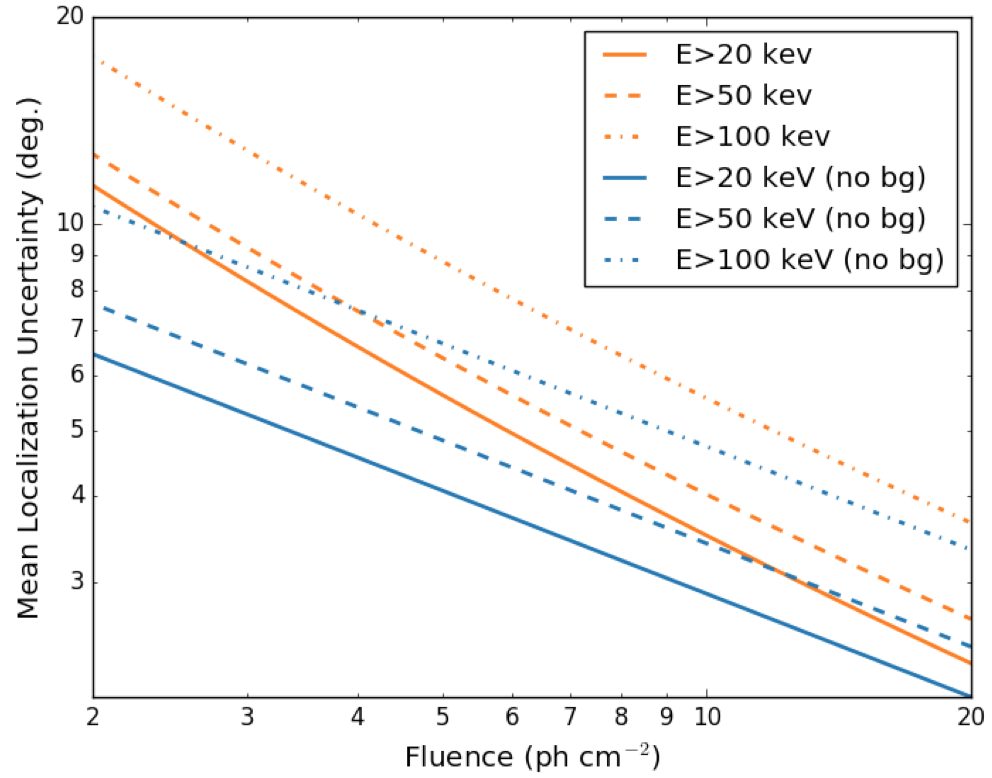}\includegraphics[height=0.255\textheight]{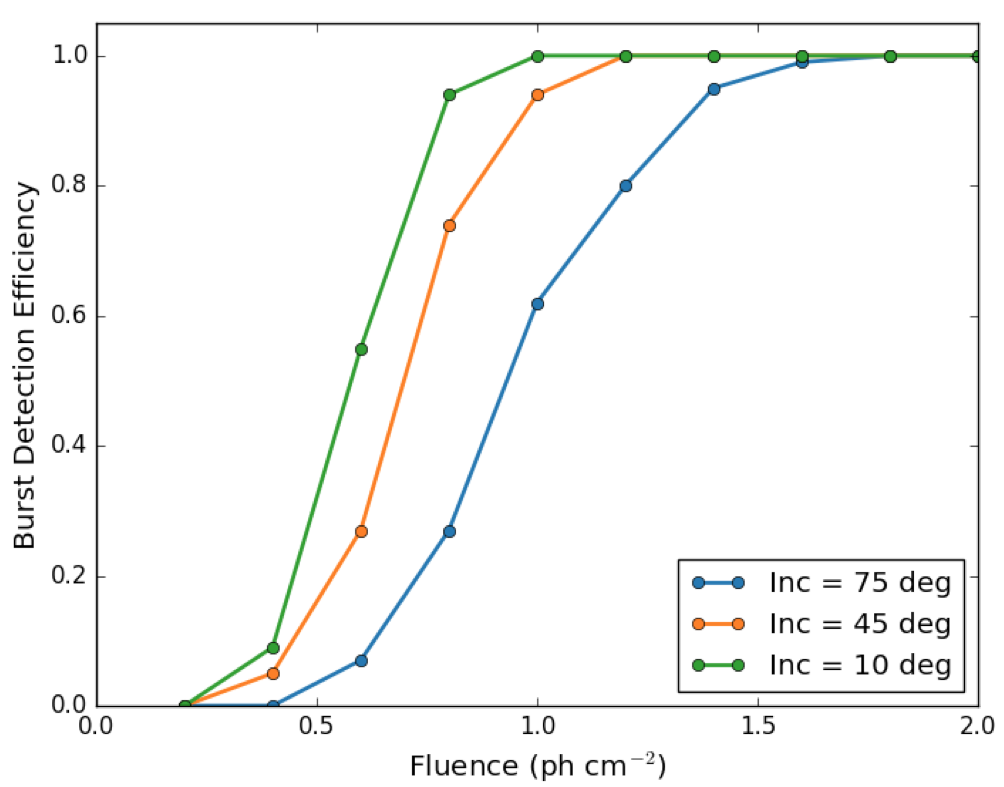}
     \caption{(Left) Localization performance for three energy thresholds both with and without estimated background.  At larger fluences, the background is sub-dominant and the performance approaches the signal-limited case.  (Right) The detection efficiency at three incidence angles and a 20 keV threshold.
}
     \label{fig4}
\end{figure}

\section{Conclusion}
\vspace{-0.15in}

Because of the rarity of nearby NS-binary merger events, and the expected short-durations of their gamma-ray counterparts ($<$2 s), only the deployment of new, low-cost instruments will provide the necessary full-sky gamma-ray coverage that will enable these SGRB detections. The addition of MERGR, deployed by 2021, will single-handedly increase the community's observing capability for nearby SGRBs to provide nearly full-sky coverage together with existing instruments (Figure~5). The compact binary mergers that generate electromagnetic counterparts are at the edge of sensitivity for the advanced interferometry observatories. All-sky gamma-ray coverage enabling an increase in the number of detections of SGRBs will be invaluable in correlating with ground-based gravitational wave observations in search of sub-threshold events. We argue that the rapid deployment of instruments like MERGR are of immediate urgency to the multi-messenger astrophysics community.

\begin{figure}[t]
\begin{minipage}[c]{0.61\textwidth}
\includegraphics[width=\textwidth]{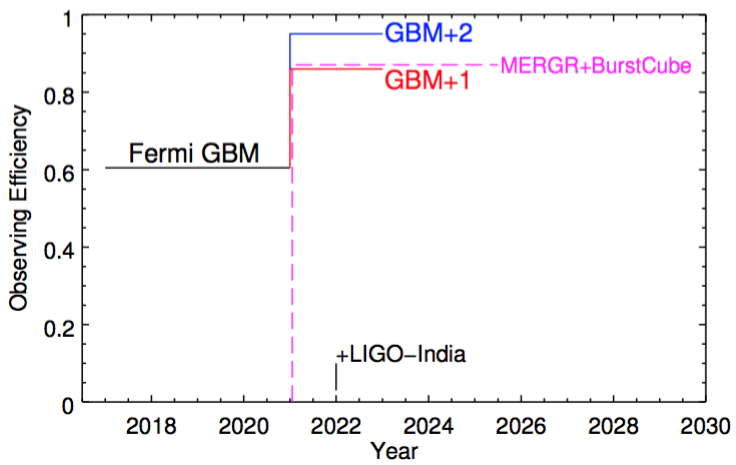}
\end{minipage}\hfill
\begin{minipage}[c]{0.36\textwidth}
  \caption{In concert with Fermi GBM, an additional large-FOV gamma-ray instrument will increase the observing efficiency (i.e. fractional sky coverage) from $\sim60\%$ (black) to $\sim85\%$ (red), with $40-45\%$ overlap in sky coverage. Addition of a third instrument (blue) will further boost observing efficiency to $\sim95\%$ of the sky with $\sim65-70\%$ overlap. Joint MERGR + BurstCube \cite{rac17} efficiency in the absence of Fermi GBM is indicated in magenta.
}
     \label{fig5}
\end{minipage}
%\vspace{-4mm}
\end{figure}

\end{document}